\documentclass[12pt]{article}
\usepackage{epsf}
\usepackage[utf8]{inputenc}
\usepackage[english]{babel}

\begin{document}

\title{
{\bf QCD and nuclear physics. How to explain the coincidence between the action radius of nuclear forces and the characteristic scale of the neutron-neutron 
electrostatic interaction?}}
\author{A.A. Godizov\thanks{E-mail: anton.godizov@gmail.com}\\
{\small {\it SRC Institute for High Energy Physics of NRC ``Kurchatov Institute'', 142281 Protvino, Russia}}}
\date{}
\maketitle

\vskip-1.0cm

\begin{abstract}
An attempt is made to interpret, in the framework of QCD, the likeness between the shapes of the neutron-neutron strong and electrostatic interactions at the distances 
$\sim 1\div 3$ fm.
\end{abstract}

\section*{1. Introduction}

The discovery of the lightest vector charmonium $J/\psi$ in 1974 and the explanation of its decay width narrowness in the framework of perturbative quantum chromodynamics 
(QCD) \cite{yndurain} induced a drastic rise of interest in this quantum-field model from the scientific community, and for more than 30 years QCD is the only world-wide 
recognized candidate for the position of fundamental theory of strong interaction. At present, a variety of very effective theoretical approximations exists: chiral 
perturbation theory, lattice QCD, the Skyrme model, relativistic mean field theory, the QCD sum rules, nonrelativistic models with ``realistic'' potentials, contact 
interaction models, bag models, {\it etc.} (The modern state of the low-energy nuclear interaction theory is expounded in pedagogical review \cite{review}.) However, in 
spite of the impressive successes in development of these approaches, many important problems have not been solved yet, concerning the achievement of agreement between 
experiment and the QCD outcomes.

In this eprint an attempt is made to get within QCD an answer to the question from the title. Certainly, we will indispensably touch upon the problem of accordance between 
dynamics of low-energy nuclear systems and the global color structure of QCD.

\section*{2. Comparison of the two characteristic scales}

The available experimental data on the neutron charge form factor allow, in principle, a straight model-independent restoration of the shape of the electrostatic interaction 
between neutrons, but, for simplicity, we use some test parametrization. 

\begin{figure}[ht]
\vskip -0.3cm
\hskip 0.1cm
\epsfxsize=7.7cm\epsfysize=7.7cm\epsffile{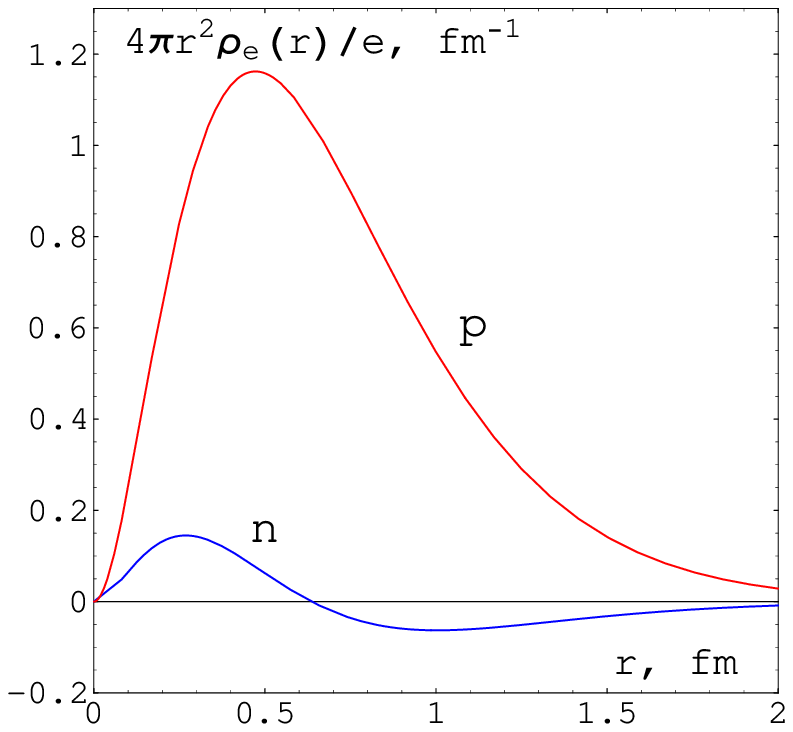}
\vskip -7.85cm
\hskip 9cm
\epsfxsize=7.85cm\epsfysize=7.85cm\epsffile{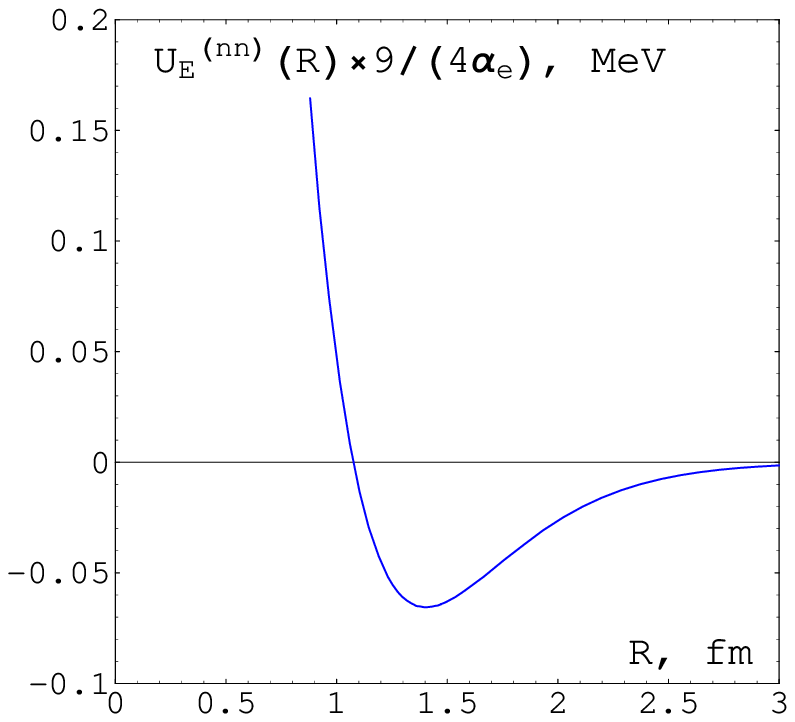}
\vskip -0.2cm
\caption{Approximate radial distributions of electric charge in proton and neutron (on the left) and the potential energy $U^{(nn)}_E(R)$ of the neutron-neutron 
electrostatic interaction as function of the distance $R$ between their centers of mass (on the right).}
\label{zarexp}
\end{figure}

The electric charge distribution in nucleon can be represented as the superposition of two functions which determine the distributions of the quarks $u$ and $d$. Let us 
approximate the charge density inside proton by $\rho_e^{(p)}(r)=\frac{4e}{3}\exp(-r/b)/(8\pi b^3)-\frac{e}{3}\exp(-r/a)/(8\pi a^3)$ and, correspondingly, the charge density 
inside neutron by $\rho_e^{(n)}(r)=\frac{2e}{3}\exp(-r/a)/(8\pi a^3)-\frac{2e}{3}\exp(-r/b)/(8\pi b^3)$, where $a=$ 0.2 fm and $b=$ 0.225 fm. Such a description of the 
charge structure of nucleons seems acceptable, since in the range $0<|\vec q\,|c<$ 1 GeV the resulting charge form factor of proton, 
$F^{(p)}(\vec q\,^2)=\frac{4}{3(1+b^2\vec q\,^2/\hbar^2)^2}-\frac{1}{3(1+a^2\vec q\,^2/\hbar^2)^2}$, almost coincides with the phenomenological approximation 
$F^{(p)}_{phen}(\vec q\,^2)=\left(1+\frac{\vec q\,^2c^2}{0.71\,GeV^2}\right)^{-2}$ usually exploited in literature, and the test charge form factor of neutron, 
$F^{(n)}(\vec q\,^2)=\frac{2}{3(1+a^2\vec q\,^2/\hbar^2)^2}-\frac{2}{3(1+b^2\vec q\,^2/\hbar^2)^2}$, is consistent with the experimental outcomes (see, for instance, 
Fig. 13 in \cite{platchkov}). The corresponding radial distributions of electric charge in nucleons are presented in Fig. \ref{zarexp}, on the left. 

Having got the information on the electric charge radial distribution in neutron, we are able to calculate, in the leading approximation, the potential energy 
$U^{(nn)}_E(R)$ of the neutron-neutron electrostatic interaction as function of the distance $R$ between their centers of mass (Fig. \ref{zarexp}, on the right). The 
position of this function's minimum, $R_{min}\approx$ 1.4 fm, is not noticeably different from the action radius of nuclear forces. Namely, one can speak about the 
coincidence of these two scales characteristic for neutron and, in general, about the extraordinary similarity of the shapes of the neutron-neutron strong and electrostatic 
potentials at the distances $\sim 1\div 3$ fm.

The very fact of such a likeness is highly nontrivial, because the structure of these two interactions is, admittedly, quite different. Let us try to explain this effect 
in terms of chromodynamics.

\section*{3. The state vector of nucleon and the general structure of nucleon-nucleon interaction}

In quantum theory, nucleon is described by means of the state vector $|\psi>_{nucleon}$ which is understood as the ground configuration of three interacting fermions (the 
constituent quarks) possessing some electric charge, color, and flavor. Namely, the proton state vector is usually represented in the following form \cite{halzen}: 
$$
|p\uparrow>\;=|\psi>_{color}\otimes\;|\psi\uparrow>_{spin,flavor}\otimes\;|\psi>_{space}\;=
$$
\begin{equation}
=\frac{1}{\sqrt{6}}\,(\,|1,2,3>_c+\;|2,3,1>_c+\;|3,1,2>_c-\;|2,1,3>_c-\;|1,3,2>_c-\;|3,2,1>_c)\,\otimes
\label{state0}
\end{equation}
$$
\otimes\;\frac{1}{\sqrt{18}}\left[\,2\,(u\uparrow u\uparrow d\downarrow) - (u\uparrow u\downarrow + u\downarrow u\uparrow)d\uparrow 
+\;permutations\,\right]\otimes|\psi>_{space}\,,
$$
where $|\psi>_{space}$ denotes the coordinate part of the state vector. For neutron, we need just to change $u\leftrightarrow d$.

Some evident arguments can be put forward in favor of such a description: 
\begin{itemize}
\item the binding requirement of the wave function strict antisymmetry characteristic for systems of identical fermions (Fermi-Dirac statistics);
\item the presumption of the isotopic symmetry ($u$-quark and $d$-quark are considered as two states of the same particle with 
different values of the isospin projection $I_3$);
\item the requirement of the nucleon color singletness (the confinement);
\item the compatibility of $|\psi>_{spin,flavor}$ in (\ref{state0}) with the empirical relation $\mu^{(n)}/\mu^{(p)}\approx -2/3$ for the magnetic moments of nucleons.
\end{itemize}

First of all, we should note that, in the framework of the Standard Model, quarks $u$ and $d$ are not identical. The isospin symmetry is an effective approximate symmetry 
of strong interaction, and not a fundamental symmetry of QCD. As well, the confinement is an observable phenomenon and not a primordial attribute of the QCD Lagrangian. 
Therefore, the requirement of {\it a priori} antisymmetry of the nucleon wave function, regarding the permutations of the $u$-quark and $d$-quark quantum numbers, is 
redundant, since it is not a strict consequence of the Pauli principle (contrary to the case of $\Delta^{++}$ or $\Delta^-$).

Let us formulate the hypothesis which underlies the further reasoning.

{\bf Basic hypothesis 1.}
\begin{itemize}
\item In quantum chromodynamics, there exists a gauge (accomplished, if necessary, by global $SU(3)_c$ transformation) and a renormalization scheme 
consistent with this gauge, wherein the state vector of isolated nucleon at rest has the following structure, in the leading approximation:
\begin{equation}
|\psi\uparrow>_{nucleon}\;=|\psi>_{flavor}\otimes\;|\psi>_{color}\otimes\;|\psi\uparrow>_{spin}\otimes\;|\psi>_{space}\;=
\label{state1}
\end{equation}
$$
=|\psi>_{flavor}\,\otimes\;\frac{1}{\sqrt{2}}\,(\,|1,2,3>_c-\;|2,1,3>_c)\,\otimes\,
\frac{1}{\sqrt{6}}\left[\,2\,(\uparrow \uparrow \downarrow) - (\uparrow \downarrow + \downarrow \uparrow)\uparrow\right]\,\otimes\,|\psi>_{space}\;,
$$
where $|\psi>_{flavor}\;=|\,u\,u\,d>$ for proton and $|\psi>_{flavor}\;=|\,d\,d\,u>$ for neutron.
\end{itemize}
Note, that such a structure of the nucleon state vector is in agreement with both the Pauli principle and the observed relation $\mu^{(n)}/\mu^{(p)}\approx -2/3$.

State (\ref{state1}) is not a color singlet. Consequently, the problems of the gluon confinement and of the color degeneracy of hadron states (which is not observed) 
emerge. These problems are discussed below, but at this point we would like to emphasize the validity of the following equalities for the state vector (\ref{state1}):
\begin{equation}
<\psi|\hat t^{(i)}_a|\psi>\,=0\;\;(a=1,...\,,7;\;i=1,2,3)\,,
\label{zero}
\end{equation}
$$
<\psi|\hat t^{(1,2)}_8|\psi>\,=\frac{1}{2\sqrt{3}}\;\,,\;\;\;\;\;\;<\psi|\hat t^{(3)}_8|\psi>\,=-\frac{1}{\sqrt{3}}\;\,,\;\;\;\;\;\;
<\psi|\,\hat t^{(1)}_8+\hat t^{(2)}_8+\hat t^{(3)}_8|\psi>\,=0\;\,,
$$
where $\hat t_a\equiv\lambda_a/2$ are the generators of the $SU(3)$ group fundamental representation ($\lambda_a$ are the Gell-Mann matrices) and the superscript denotes 
quark's number. 

Relations (\ref{zero}) have several important consequences:
\begin{itemize}
\item nucleons are sources of the gluon octet 8th component only; 
\item under exchanges by separate gluons, nucleons {\it do not change} their color structure at all, owing to the matrix $\lambda_8$ diagonality;
\item in the tree level, interaction of nucleons is reduced to the superposition of the constituent quark Coulomb interactions with the only distinction that 
the effective charge of the $u$-quarks in proton and of the $d$-quarks in neutron is equal to $-\frac{g_s}{2\sqrt{3}}$, and the effective charge of the $d$-quark in proton 
and of the $u$-quark in neutron is equal to $+\frac{g_s}{\sqrt{3}}$, where $g_s$ is the elementary charge of QCD, {\it i.e.}, the 
distribution of the nuclear interaction effective charge in nucleon is proportional to the electric charge distribution in neutron. 
\end{itemize}

In other words, the following physical pattern takes place. The operator of the total energy of nucleon-nucleon interaction can be represented as 
$$
\hat U = \sum_{i,j}\hat U^{gl}_{ij}+\hat U^r\,,
$$
where $\hat U^{gl}_{ij}$ are the weakly connected parts of the effective operators of the pair interaction between those constituent quarks which are contained in different 
nucleons, and $\hat U^r$ denotes other contributions (meson exchanges). Under the term ``weakly connected part of quark pair interaction operator'' we understand the 
appropriate complete sum of the weakly connected Green functions composed of bare propagators and 3-point and 2-point full strongly connected Green functions only. If the 
isolated nucleon state vector takes form (\ref{state0}), then, in view of the structure of generators $\hat t_a$, the contributions of the weakly connected parts of the 
constituent quark pair interactions into the potential energy of the nuclear system are equal to zero (the confinement case). If the isolated nucleon state vector takes 
form (\ref{state1}), then the weakly connected part of nucleon-nucleon interaction can be effectively represented as 
\begin{equation}
\hat U^{gl} = \sum_{i,j}\hat U^{gl}_{ij} = \frac{1}{12}\sum_{i,j}Q^{[1]}_iQ^{[2]}_j\left[\hat U^{CS}_{ij}+\hat U^{CM}_{ij}\right]\,,
\label{nerpot}
\end{equation}
where $Q=-1$ for the $u$-quarks in proton and for the $d$-quarks in neutron, and $Q=+2$ for the $d$-quark in proton and for the $u$-quark in neutron. The structure of the 
operators $\hat U^{CS}_{ij}$ and $\hat U^{CM}_{ij}$ {\it does not depend} on the quark colors and, in the leading approximation, on their flavors. Operators 
$\hat U^{CS}_{ij}$ are related to the effective chromostatic (central) potential $V^{CS}(r)$ which includes the influence of the QCD vacuum polarization. Operators 
$\hat U^{CM}_{ij}$ are related to the effective potential $V^{CM}(\vec r\,,\vec n_1,\vec n_2)$ of quark-quark chromomagnetic (nonscalar) interaction, the unit vectors 
$\vec n_1$ and $\vec n_2$ determine the quark spin orientations.

For the phenomenological description of nucleons as the 3-fermion systems with the effective charge structure $\{-1,-1,+2\}$, the dominance of the weakly connected parts of 
the quark pair interactions over other contributions is necessary. Therefore, in addition to basic hypothesis 1, we need

{\bf Basic hypothesis 2.}
\begin{itemize}
\item In the systems of low-energy nucleons, the interaction of nucleons at the distances higher than 0.5 fm is dominated by the weakly connected parts of the effective 
quark-quark interactions.
\end{itemize}
Note, that although the presumed effective structure (\ref{state1}) of the nucleon state vector is related to some concrete renormalization scheme associated with a certain 
gauge, the content of the basic hypothesis 2 itself is not related to any gauge or renormalization scheme.

Under such an approach, the short-rangeness of the chromostatic interaction of nucleons is conditioned, mainly, by their effective integral neutrality (like the 
short-rangeness of interaction between helium atoms), in contrast to the models with pion and other meson exchanges wherein the short-rangeness is related to the meson 
field massiveness (certainly, basic hypothesis 2 implies not the absence of meson exchanges, but their subdominance in the low-energy nuclear interaction 
regime). Besides, at the distances $r\gg$ 1 fm the effective potential $V^{CM}(\vec r,\vec n_1,\vec n_2)$ should decrease significantly faster than $r^{-3}$, for the faster 
drop of the nonscreened chromomagnetic interaction in comparison with the magnetic interaction of the quarks.

\section*{4. Discussion}

\subsection*{4.1. Qualitative form of effective nuclear potential}

\begin{figure}[ht]
\begin{center}
\epsfxsize=7.7cm\epsfysize=7.7cm\epsffile{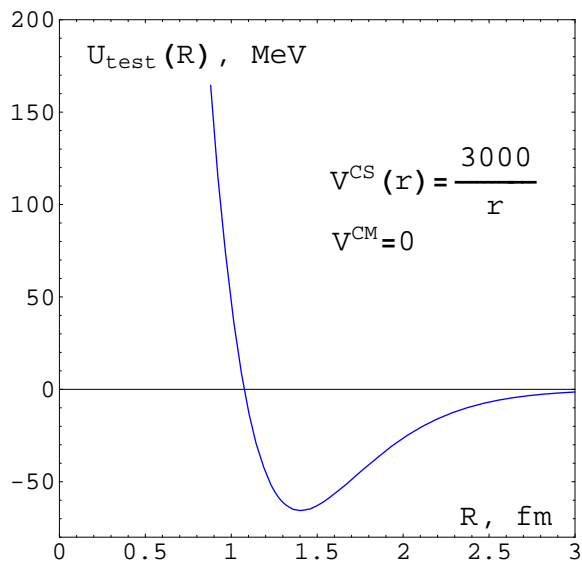}
\end{center}
\caption{The nucleon-nucleon interaction energy as function of the distance $R$ between their centers of mass for the test case of quark-quark interaction 
$\{V^{CS}(r)=3000/r$, $V^{CM}=0\}$.}
\label{potest}
\end{figure}

The constituent quark distribution densities can be extracted from the proton and neutron experimental form factors. Then, in accordance with the aforesaid, the 
potential energy of nucleon-nucleon interaction could be introduced as the sum of the energies of the constituent quark pair interactions. In Fig. \ref{potest} the result 
is presented for the test case $\{V^{CS}(r)=3000/r$, $V^{CM}(\vec r\,,\vec n_1,\vec n_2)=0\}$ (of course, the true effective potential of quark-quark 
chromostatic interaction could essentially differ from the Coulomb one, and the true chromomagnetic potential is nonzero).

\subsection*{4.2. The range of applicability and the color degeneracy problem}

Basic hypothesis 1 states that structure (\ref{state1}) corresponds to isolated nucleon at rest. Therefore, the adequacy of the considered approximation is expected for the 
systems of low-energy nucleons only. The color structure of state vector (\ref{state1}) could, in principle, take place for the hyperons $\Sigma$ and $\Xi$ of the lightest 
baryon octet. But it is absolutely unsuitable, say, for the decuplet of the light spin-3/2 baryons which are color singlets.

Nonetheless, as nucleon is not a color singlet in the proposed approach, so we should comment on the observed absence of the color degeneracy. The nucleon color 
non-singletness itself is not a principal problem, since, in fact, {\it in experiment we observe not the color singletness of hadron states, but the absence of their 
degeneracy in color}. The latter phenomenon could be explained not only by the local colorlessness of all the hadrons, but, alternatively, by the spontaneous or dynamical 
breaking of color symmetry for the low-mass states (say, $SU(3)\to SU(2)$\footnote{The simplest variant of spontaneous breaking of $SU(3)_c$ gauge symmetry requires 
introduction of the following gauge-invariant terms into the QCD Lagrangian: ${\cal L}^* = 
\left|\left(\delta_{jk}\partial_{\mu}-i\,g_s\frac{\lambda_{jk}^a}{2}B_{\mu}^a\right)\Phi_k\right|^2-\frac{\kappa^2}{2}\left[\Phi_k^+\Phi_k-\frac{\omega^2}{2}\right]^2$, 
where $B_{\mu}^a$ is the gluon octet, $\lambda^a$ are the Gell-Mann matrices, $g_s$ is the strong coupling constant, and $\Phi_k$ is a triplet of complex scalar fields. As a 
result, three components of the gluon octet remain massless, others acquire effective masses. These masses should be $\ll m_{\pi^0}$, so that all the gluon octet components 
could be treated as massless at the energies higher the lowest threshold. Besides, such a Higgs mechanism implies the emergence of some massive scalar particle (colored and 
electrically neutral). As contrasted to gluons, the mass of this particle is expected to be very high, $\gg$ 1 TeV. Otherwise, it would have already been discovered.} or 
$SU(3)\to SU(2)\otimes U(1)$), which is not forbidden by general principles. Certainly, the symmetry is restored at higher energies.

\subsection*{4.3. The isospin symmetry of nuclear forces}

Owing to the equivalence of the proton and neutron color structures, the effective interaction of nucleons possesses the isospin symmetry, in the leading approximation. 
However, we would like to point out that the color of the $u$-quark in neutron is fixed and is different from the colors of the $u$-quarks in proton. The same can be said 
about the $d$-quarks. Therefore, the exchange contributions are absent in the interaction between proton and neutron (they are distinguishable entirely), in contrast to 
interaction of identical nucleons. But, due to the rather weak overlap of the wave functions of low-energy nucleons, the exchange interaction turns out to be just a small 
corrective effect which violates the isospin symmetry on the level with the quark-quark electromagnetic interaction and the difference between $m_u$ and $m_d$.

\subsection*{4.4. The quark confinement}

As the expected range of the considered approximation validity is restricted by low-energy nuclear systems, so the foregoing reasoning is not in contradiction with the 
quark confinement hypothesis.

\subsection*{4.5. On the experimental non-observation of gluons}

In the proposed approach, nucleons are gluon field sources. Hence, one could expect that nuclear collisions are accompanied by radiation of real gluons. These gluons can 
interact between themselves (including the color exchanges). As well, they are able to annihilate into photons or to be absorbed by nucleons, but, in view of relations 
(\ref{zero}), only the 8th component of the gluon octet interacts with nucleons\footnote{Contrary to the interaction between quarks contained in different nucleons and 
between nonidentical quarks of the same nucleon, an additional attractive interaction emerges between identical quarks of the same nucleon. This interaction is related to 
the first three (massless) components of the gluon octet. It is so strong that the diquark composed of the identical quarks is less massive than the third quark (the 
$u$-quark in neutron is distributed closer to the center of mass than the $dd$-diquark).}. Let us remind, that, under such interaction, the structure of the color part of 
the nucleon state vector (\ref{state1}) keeps unchanged, {\it i.e.}, the color exchange between nucleons {\it does not occur}. Certainly, such a radiation will take place 
only if the gluon octet 8th component is massless or if its mass (acquired dynamically or due to the spontaneous breaking of color symmetry) is low enough. If this mass is 
higher than the collision energy, then gluons are not radiated.

The density of effective nuclear charge in nucleon is proportional to the electric charge density in neutron. Consequently, if the gluon octet 8th component is massless, 
then the energy spectra of the gluons radiated in nonrelativistic nuclear collisions are alike to the energy spectra of the photons radiated in neutron-neutron scattering. 
Significant divergence (about several orders of magnitude) is in the multiplicities only, due to the difference between the elementary charge values of chromodynamics and 
electrodynamics. It is quite possible that such a ``disguise'' of massless gluons as soft photons is the main reason why they are not observed in low-energy nuclear 
reactions (especially, in the reactions with participation of protons). However, gluons do not interact with leptons. Therefore, a principal feasibility exists to 
distinguish them from photons.

Concerning the non-observation of gluons even at higher energies (up to the energies characteristic for quark-gluon plasma), one should pay attention to the fact that 
strong interaction is several orders stronger than electromagnetic one. As, in addition, color can be transferred by gluons of arbitrarily small energy, so the gluon 
confinement hypothesis does not prevent numerous discussions on the gluon jet production in high-energy collisions.

\section*{5. Conclusions}

Thus, an answer to the question from the eprint title can be given in terms of the global color structure of QCD. The attractivity of such an explanation is determined as 
by its simplicity, so by the fact that the first and second basic hypotheses do not introduce any notions extraneous for QCD. Namely, these hypotheses just presume the 
dominance (for the systems of low-energy nucleons) of certain contributions which have an unambiguous interpretation in the framework of QCD. Nevertheless, only explicit 
comparison of the detailed modelling outcomes with the experimental data on the nonrelativistic scattering phase shifts and on the light nuclei spectra will allow to 
confirm or disprove the adequacy and practical usefulness of the proposed approach to description of nuclear forces which requires to introduce the specific effective 
charge structure of nucleon: $\{-1,-1,+2\}$.

\section*{Acknowledgements}

The author is grateful to his colleagues from the IHEP Division of Theoretical Physics and from CERN for their questions and numerous remarks which helped to improve the 
formulations.

\end{document}